\documentclass{emulateapj}

\shorttitle{Nonthermal Emission in Cas A}

\shortauthors{Patnaude et al.}

\begin{document}

\title{A Decline in the Nonthermal X-ray Emission from Cassiopeia A}

\author{Daniel J.~Patnaude\altaffilmark{1}, 
Jacco Vink\altaffilmark{2}, J.~Martin Laming\altaffilmark{3}, and
Robert A.~Fesen\altaffilmark{4}}
\altaffiltext{1}{Smithsonian Astrophysical Observatory, Cambridge, 
MA 02138, USA}

\altaffiltext{2}{Astronomical Institute, Utrecht University, P.~O. Box 80000,
3508 TA Utrecht, The Netherlands}

\altaffiltext{3}{Space Science Division, Naval Research Laboratory, Code 
7674L, Washington, DC 20375, USA}

\altaffiltext{4}{6127 Wilder Lab, Department of Physics \& Astronomy, 
Dartmouth College, Hanover, NH 03755, USA}

\begin{abstract}

We present new {\it Chandra} ACIS-S3 observations of Cassiopeia A which, when
combined with earlier ACIS-S3 observations, show evidence for a steady $\sim$
1.5--2\% yr$^{-1}$ decline in the 4.2--6.0 keV X-ray emission between the years
2000 and 2010. The computed flux from exposure corrected images over the entire
remnant showed a 17\% decline over the entire remnant and a slightly larger
(21\%) decline from regions along the remnant's western limb. Spectral fits of
the 4.2-6.0 keV emission across the entire remnant, forward shock filaments,
and interior filaments indicate the remnant's nonthermal spectral powerlaw
index has steepened by about 10\%, with interior filaments having steeper
powerlaw indices. Since TeV electrons, which give rise to the observed X-ray
synchrotron emission, are associated with the exponential cutoff portion of the
electron distribution function, we have related our results to a change in the
cutoff energy and conclude that the observed decline and steepening 
of the nonthermal X-ray emission is consistent with a deceleration of the
remnant's $\simeq$5000 km s$^{-1}$ forward shock of $\approx$ 10--40 km s$^{-1}$
yr$^{-1}$.  

\end{abstract}


\section{Introduction}

Supernova remnants (SNRs) have long been considered to be the primary source of
Galactic cosmic-rays (CRs) below the {\it knee} of the cosmic-ray spectrum,
$\sim$ 10$^{15}$ eV. TeV $\gamma$-ray observations of SNRs such as RX
J1713.7-3946 and RX J0852.0-4622 provide evidence for the acceleration of ions
\citep{aharonian07a,aharonian07b}. However the TeV emission can also be
attributed to inverse-Compton scattering by the same electron population that
produces the X-ray synchrotron emission.

Viewed in X-rays, the young ($\sim$ 330 yr; \citealt{fesen06}) Galactic SNR
Cassiopeia A (Cas A) consists of a shell whose emission is dominated by
emission lines from O, Si, S, and Fe
\citep[e.g.,][]{vink96,hughes00,will02,will03,hwang03,laming03}.  Exterior to
this shell are faint X-ray filaments which mark the location of the forward
shock. The emission found here is nonthermal X-ray synchrotron emission from
shock accelerated electrons \citep{allen97,gotthelf01,vink03}.  These forward
shock filaments are observed to expand with a velocity of $\simeq$ 5000 km
s$^{-1}$ \citep{delaney03,patnaude09}, assuming a SNR distance of 3.4 kpc
\citep{reed95}. 

Nonthermal emission filaments are also observed in the interior of the SNR and
are believed to be either forward shock filaments seen in projection
\citep{delaney04,patnaude09} or associated with efficient acceleration of
electrons at the SNR reverse shock \citep{uchiyama08,helder08}.  Fluctuations
in both exterior and interior nonthermal filaments have also been reported
\citep{patnaude07,uchiyama08,patnaude09}, and the variability is cited as
evidence for rapid synchrotron cooling of TeV electrons in mG--scale fields. A
two to four year timescale for variations is evidence for efficient diffusive
shock acceleration in SNR shocks, or alternatively the variations are seen as
evidence for magnetic field fluctuations due to plasma waves behind the shock
\citep{bykov08}.

Emission from Cas A has been seen at energies up to $\sim$ 40 keV with the {\it
Suzaku} HXD PIN detector \citep{maeda09}, up to 100 keV with {\it CGRO} OSSE
and {\it Integral} IBIS \citep{the96,renaud06}, and 
GeV emission has been detected using {\it
Fermi}--LAT \citep{abdo10}.  The {\it Fermi} observations do not rule out
either a leptonic origin to the GeV emission from a combination of nonthermal
bremsstrahlung and inverse--Compton emission or a hadronic origin from neutral
pion decay. Finally, Cas A has been detected at even higher TeV energies with
HEGRA, MAGIC and Veritas \citep{aharonian01,albert07,humensky08}.
Interestingly, the centroids for the GeV--TeV emission are located in the
western region of Cas A, where the nonthermal X-ray emission is brightest
\citep{helder08,maeda09}. 

Here we present {\it Chandra} ACIS-S3 observations of Cas A taken in 2009 and 2010 which,
when compared to ACIS-S3 observations taken between 2000 and
2007, show the remnant's nonthermal X-ray emission in the 4.2--6.0 keV band
to have decreased at a rate of $\simeq$ 1.5--2.0\% yr$^{-1}$. 
In \S~2 we discuss our observations, data reduction, and spectral analysis and 
in \S~3 we discuss our results and offer some conclusions about the current
and future evolution of the nonthermal emission in Cas A.

\section{Observations, Data Reduction, and Analysis}

\begin{figure}
\plotone{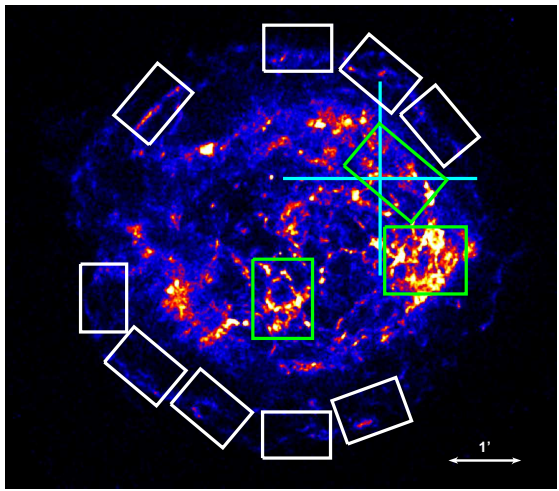}
\caption{Exposure corrected image of Cas A between 4.2--6.0 keV. The
white boxes mark approximately the locations of the spectra 
extracted from the forward shock, while the green boxes mark those
regions where spectra were extracted from the interior filaments.
The cyan cross marks the location and 68\% confidence limit 
of the {\it Fermi} centroid.}
\label{fig:casa10}
\end{figure}

Cas A has been observed extensively with {\it Chandra} and we have made use of
several GO observations taken between 2000 and 2010, including the 2004 VLP
(PI: Hwang). We reprocessed each epoch dataset listed in Table~\ref{tab:data}
using CIAO\footnote{\url{http://chandra.harvard.edu}} version $4.2$ and CalDB
$4.2.2$. The more recent observations taken in 2007, 2009, and 2010 were split
due to spacecraft thermal constraints. We merged these split observations
into a single event lists for each epoch.

We filtered the event lists for those events with energies between 4.2--6.0 keV
and performed an exposure correction assuming a monochromatic 5.1 keV source.
The exposure corrected image is in units of photons cm$^{-2}$ s$^{-1}$
pix$^{-1}$, so to compute the flux of 5.1 keV photons cm$^{-2}$ s$^{-1}$, we
integrate over the number of pixels. The latest dataset, taken in November
2010, is shown in Figure~\ref{fig:casa10}. 

\begin{figure}
\plotone{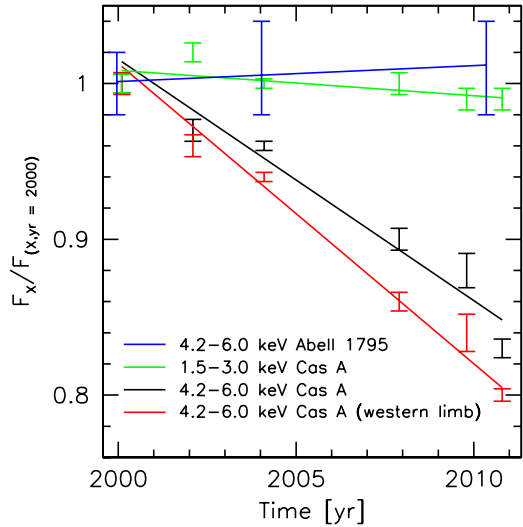}
\caption{Comparison of 4.2--6.0 keV flux in Cas A compared
to the year 2000 observations. The black curve and data correspond to
changes in the whole SNR, while the red curve and data correspond
to changes in the western portion of Cas A only. For reference we also
show the 1.5--3.0 keV flux from Cas A (fluxed at 1.85 keV) as well
as the 4.2--6.0 keV emission from the cluster Abell 1795. The observed
decline in the 4.2--6.0 keV emission in Cas A corresponds to a fractional
decline of -(1.5$\pm$0.17)\% yr$^{-1}$ across the whole SNR, and 
-(1.9$\pm$0.10)\% yr$^{-1}$ in the western limb.}
\label{fig:decline}
\end{figure}

As seen in the black curve in Figure~\ref{fig:decline}, there is a clear
decrease in the overall 4.2--6.0 keV emission from Cas A. This decline does not
appear to be an instrumental artifact. Several sources have been observed as
{\it Chandra} calibration targets, including the galaxy cluster Abell~1795. In
Figure~\ref{fig:decline}, we also plot the 4.2--6.0 keV emission from
Abell~1795, showing that the emission from this cluster has not varied by more
than 1--2\% over $\sim$ 10 years (Vikhlinin 2010, private communication).  As a
further check on whether our result for Cas A could be due to an instrumental
or calibration artifact, we plot in Figure~\ref{fig:decline} the 1.5--3.0 keV
emission (exposure corrected at 1.85 keV) and find that emission in that band
has declined by less than 1\% in 11 years.  In addition, the Galactic SNR,
G21.5--0.9, has been observed extensively with {\it Chandra} ACIS-S3 as a
steady 2.0--8.0 keV continuum source, and changes in the flux from that source
are not observed at the level which we report here \citep{tsujimoto10}.
Finally, we note that \citet{katsuda10} reported no change in the X-ray
synchrotron emission from SN~1006, and \citet{heinke10} reported a decline in
the temperature of the central point source in Cas A, and showed that the
ACIS-S3 efficiency in the hard band is stable.

As an additional test of this flux decline, we fitted the 4.2--6.0 keV
continuum emission in Cas A at each epoch to a powerlaw model in {\sc
XSPEC\footnote{\url{http://heasarc.gsfc.nasa.gov/docs/xanadu/xspec/}}} version
$12.6$. We find that the modeled flux does decrease with time in this energy
band, consistent with our analysis of the exposure corrected images. As shown
in Column~2 of Table~\ref{tab:data}, the fitted spectrum appears to steepen
with time.

The results of our spectral fits listed in Column 2 of Table~\ref{tab:data} are
across the entire SNR and thus can include emission from electrons accelerated
at the remnant's forward shock, the reverse shock, or at the contact interface.
To determine whether the changing emission is from the outer forward shock
filaments, interior regions or possibly both, we extracted spectra from several
regions marked with boxes in Figure~\ref{fig:casa10} in all datasets and again
fit the nonthermal emission to a powerlaw model. Galactic absorption has a
negligible effect on 5.0 keV photons, so we did not model it here. 


In Columns 4--6 of Table~\ref{tab:data}, we list the results from the spectral
fits for the forward shock regions (the white boxes in
Figure~\ref{fig:casa10}), the interior regions (green boxes in
Figure~\ref{fig:casa10}), and then also just the interior regions that coincide
with the {\it Fermi} centroid \citep{abdo10}.  As can be seen in
Table~\ref{tab:data}, the spectral shape of the forward shock filaments varies
with time, with $\Gamma_{\mathrm{FS}}$ $\sim$ 2.3--2.6 between $2000.1$ and
$2010.8$. The interior filaments and the region associated only with the {\it
Fermi} centroid in Figure~\ref{fig:casa10} also show an increase in the
powerlaw index over the same time period.  Interior filaments show a steeper
spectral shape than that seen for forward shock regions. We also examined
changes in intensity in the western portion of Cas A and found that the
emission from this region is also decreasing with time. This is shown as the
red curve in Figure~\ref{fig:decline} and is seen to be steeper than the
remnant's overall nonthermal emission decline rate (black curve).

\section{Discussion}

As shown in Figure~\ref{fig:decline} and listed in Table~\ref{tab:data}, we
find a decline in the 4.2--6.0 keV emission in Cas A over an approximately 11
year span. A linear least squares fit to the fractional change in 4.2--6.0 keV
emission indicates a fairly steady decline in the nonthermal emission of
(1.5$\pm$0.17)\% yr$^{-1}$ across the entire SNR. In western regions coincident
with the {\it Fermi} centroid, the rate of decline appears to be slightly
higher at (1.9$\pm$0.1)\% yr$^{-1}$.

A much slower decline in Cas A's radio emission from GeV electrons has been
known for some time \citep{shklovskii60,baars77}. \citet{reichart00} report
that Cas A has generally been fading at a rate of $\sim$ $0.6$--$0.7$\%
yr$^{-1}$ over a range of radio frequencies.  The observed decline in Cas A's
radio emission has been interpreted to be a result of the adiabatic expansion
of the supernova remnant since radio synchrotron emitting electrons have no
appreciable radiative losses \citep{shklovskii60,anderson96}.  

In contrast, the remnant's X-ray synchrotron emission is the result of
electrons accelerated within the last decade, as their radiative loss times are
of the order ten years, or possibly less. Thus, X-ray synchrotron emission is
much more sensitive to the present day acceleration time and radiative losses.
For typical SNR parameters, synchrotron X-rays are produced in large part by
the exponential tail of the electron distribution \citep[e.g.][]{ellison05}.
Thus, any energy loss will result in a large drop in the emissivity, such as
observed here in Cas A.


In order to interpret the 4.2--6.0 keV flux changes seen in Cas A's X-ray continuum emission,
we now examine how much of this change can be attributed to a change in the spectral
properties alone. \citet{zirakashvili07} considered the nonthermal X-ray spectrum
when accounting for losses from diffusive shock acceleration and synchrotron
radiation. In that case, 
the synchrotron emission can be approximated by \citep{zirakashvili07}:

\begin{equation}
N(E) = \phi\, B^{\Gamma_0} \times E^{-\Gamma_0}\exp{\left(-\sqrt{\frac{E}{E_c}}\right)},
\label{eqn:ne}
\end{equation}
where $E$ is the energy of the emitted photon, which here we assume to 
be 5.1 keV, $E_c$ is the cutoff energy in keV, and
$\Gamma_0$ is the slope of the powerlaw well below the cutoff energy.
We have split here the normalization constant into two factors, for reasons that
we will explain below. One factor
directly corresponds to the downstream magnetic field strength 
\cite[c.f.][equation~3.31]{ginzburg65}, where $\phi$\ relates to the overall electron
acceleration efficiency.
Equation~\ref{eqn:ne} 
implies that the spectral slope $\Gamma$ in X-rays is given by \citep{vink99}:
\begin{equation}
\Gamma = \Gamma_0 + \frac{1}{2}\sqrt{\frac{E}{E_c}} \ .
\label{eqn:gam}
\end{equation}

\noindent
Thus, the measured spectral index $\Gamma$ can be related to the cutoff energy by:

\begin{equation}
E_c = \frac{E}{4(\Gamma - \Gamma_0)^2} \ .
\end{equation}

For $\Gamma_0$, we assume the value inferred from the radio spectral index, namely
$\Gamma_0 = \alpha + 1 = 1.78$ \citep{baars77}, or if nonlinear effects are important,
$\Gamma_0$ may be as low as $1.25$, corresponding to a particle spectral index
of $1.5$ \citep{malkov01}. For $\Gamma_0 = 1.5$, $E_c$= 1.0 -- 2.0 keV, 
assuming $\Gamma$ = 2.6 -- 2.3 and $E$ = 5.1 keV. Differentiating
Equation~\ref{eqn:gam} with respect to time under the assumption that only
the cutoff energy varies leads to:

\begin{equation}
\frac{dE_c}{dt} = -4 \sqrt{\frac{E_c}{E}} E_c \frac{d\Gamma}{dt} \ .\label{eqn:ecdot}
\end{equation}

\noindent
According to Table~\ref{tab:data}, we measure $d\Gamma/dt$ = 0.018 yr$^{-1}$ in
the western portion of Cas A, which
translates to $dE_c/dt$ = $-$(0.032$\pm$0.008 -- 0.090$\pm$0.015) keV yr$^{-1}$ 
for a $E_c$ = 1.0 -- 2.0 keV. Likewise, from Table~\ref{tab:data} for the forward
shock, $d\Gamma/dt$ = 0.022 yr$^{-1}$, which implies $dE_c/dt$ = $-$(0.039$\pm$0.004 -- 0.101$\pm$0.008)
keV yr$^{-1}$. 

We
can relate the cutoff energy to the shock velocity by \citep{zirakashvili07}:
\begin{equation}
E_c \approx 2.2\eta^{-1}\left(\frac{V_s}{3000 \mathrm{\ km \ s^{-1}}}\right)^2 \ \mathrm{keV,}
\end{equation}

\noindent
where $\eta$ is the diffusion parameter, and equals one in the case of 
Bohm--diffusion. Therefore,
the change in cutoff energy relates directly to a change in shock velocity as:

\begin{equation}
\frac{dV_s}{dt} =  2.0\times10^6 \left(\frac{V_s}{\mathrm{\ km \ s^{-1}}}\right)^{-1} 
\eta \frac{dE_c}{dt} \ .
\end{equation}
The forward shock of Cas A has been measured to be $\simeq$ 
5000 km s$^{-1}$ \citep{delaney03,patnaude09}, so $dV_s/dt$ = $-$(16$\pm$3 -- 40$\pm$5) 
km s$^{-1}$ yr$^{-1}$. 
The nature of the reverse shock is, however, more uncertain.
\citet{helder08} have argued that the western region of Cas A
has a velocity as high as 6000 km s$^{-1}$, 
implying that the reverse  shock in our frame is  
almost stationary. 
This could indeed be the case given a possible interaction with
a molecular cloud in the west \citep{keohane96}. However,
given the uncertainties in the nature of the interior filaments and the
shock velocity of the reverse shock we
defer a discussion on the changes in the western region of Cas A 
until more observations are completed.

So far we have only interpreted the change in spectral slope as due to a 
change in shock velocity.
In addition, using the same framework, we can also interpret the observed
flux changes.
Differentiating Equation~\ref{eqn:ne} with respect to time we obtain:
\begin{eqnarray}
\frac{dN(E)}{dt} =  & E^{-\Gamma_0}\exp\Bigl( -\sqrt{\frac{E}{E_c}}\Bigr)\times
\nonumber \\
&\Bigl[
\frac{d\phi}{dt} B^{\Gamma_0} + \phi \frac{d B^{\Gamma_0}}{dt} -
2\phi\,B^{\Gamma_0}\frac{d\Gamma}{dt}
\Bigr],\label{eqn:fluxchange0}
\end{eqnarray}
where we made use of Equation~\ref{eqn:ecdot} to rewrite the last
term. In what follows we will estimate the influence of each term separately.

To start with the last term describing the flux change due to the
evolution of cut-off energy,
it is clear from Equation~\ref{eqn:fluxchange0} that a change in $E_c$ alone
will lead to a fractional change in flux of:
\begin{equation}
\frac{1}{N(E)}\frac{dN(E)}{dt} = -2\frac{d\Gamma}{dt}.
\end{equation}
This directly connects the change in spectral slope to the related change in
flux. For the values listed above, for both the forward shock and the western
region, we expect a fractional change of $\approx$ 4\% yr$^{-1}$,
which is about twice the value that is observed.

For the other two terms in Equation~\ref{eqn:fluxchange0} we must make
some additional assumptions about their dependence on time.
Although this adds some uncertainty, it is nevertheless worthwhile
for making order of magnitude estimates.
We limit our discussion to the forward shock region as both its shock
evolution and density structure are better understood.

For the normalization factor $\phi$, we
assume that to first order it corresponds directly to the amount of gas
entering the forward shock per unit of time:
\begin{equation}
\phi \propto 4\pi r^2\rho V_s \propto t^{m-1},
\end{equation}
where we have used both the fact that the forward shock is likely to be
moving through the progenitor wind of the supernova, hence 
$\rho \propto r^{-2}$, and also the idea that the forward
shock evolution is self-similar with $r\propto t^m$. Note also
that this equation is only valid if synchrotron losses are large,
otherwise $\phi$ would be proportional to the total amount
of material swept up by the forward shock, rather than by
what is being swept up now.
Differentiating the above equation with respect to time shows that
the fractional change in the flux due to a change in $\phi$
is estimated to be:
\begin{equation}
\frac{1}{\phi}\frac{d\phi}{dt} = (m-1)t^{-1}.
\end{equation} 
For $m=0.66$ this corresponds to a fractional flux change of $-0.1$\% yr$^{-1}$,
much smaller than the flux changes due to a change in cut-off energy.

For estimating the effect of the second term in 
Equation~\ref{eqn:fluxchange0} we make the assumption
that  $B^2 \propto \rho V_s^\beta$, as suggested by  \citep{volk05},
with $\beta=2$ \citep{volk05} or $\beta=3$  \citep{vink08,bell04}.
We therefore have:
\begin{eqnarray}
B^{\Gamma_0} \propto (\rho V_s^\beta)^{\Gamma_0/2} \propto (t^{-2m}t^{(m-1)\beta})^{\Gamma_0/2} \nonumber \\
= t^{\Gamma_0[m(\beta-2)-\beta]/2},  \ 
\end{eqnarray}
for which we have again made use of the idea that the
density structure falls off as $1/r^2$ and $r\propto t^m$.

The fractional change in $B^{\Gamma_0}$, and therefore, the expected
fractional flux change is expected to be
\begin{eqnarray}
\frac{1}{B^{\Gamma_0}}\frac{d B^{\Gamma_0}}{dt} = 
(t^{-2m}t^{(m-1)\beta})^{\Gamma_0/2} \nonumber \\
= t^{\Gamma_0(m(\beta-2)-\beta)/2}  \ ,
\end{eqnarray}
corresponding to a flux change of
-0.45\% yr$^{-1}$ for 
$\beta=2$, $m=0.66$ and $\Gamma_0=1.5$ or -0.54\% yr$^{-1}$ for $\beta=3$. 
These values are larger than those due to changes in the first term ($\phi$),
but still smaller than the fractional changes predicted from changes
in the cut-off energy alone. 
It is important to note that these values are only valid for the forward shock, for which 
we have a reasonable estimate of both the preshock density and $m$. 

There are a number of caveats in the simple analysis above which may affect our
result. For instance, \citet{schure10} point out that if non-steady state
acceleration effects are considered, the exponential cutoff may have a
dependence that is different than that in Eq.~\ref{eqn:gam}.  Additionally, we
are averaging over an ensemble of shock conditions over a span of ten years;
the shock conditions and powerlaw spectra will undoubtedly vary from position
to position, and a sum of powerlaw spectra does not yield a powerlaw \citep[see
also the discussion by][]{helder08}.  In that respect, it is surprising that
the hard X-ray spectrum of Cas A is best described by a powerlaw
\citep{renaud06} with little evidence for a gradual steepening of the hard
X-ray spectrum with energy as predicted by Equation~\ref{eqn:ne}.

It is also possible that not all the emission arises from synchrotron 
emitting electrons. \citet{helder08} point out that $\ga$ 50\% of the
emission in the west is from synchrotron emission, and the remainder is likely
from thermal continuum. The loss time for thermal 
electrons is much greater than the 2--4 yr timescale seen in the TeV electrons. 
That is to say, over the
timescales we are investigating, variations in thermal continuum emission 
will be much smaller than variations in the nonthermal emission and it is possible
that the emission from the thermal component may increase over the approximately
11 yr time span of our observations. 



\section{Summary and Conclusion} Our analysis of the 4.2--6 keV flux of Cas A
shows a decline of 1.5\% yr$^{-1}$ in the nonthermal X-ray emission across the
entire SNR over 11 years, with a slightly larger decline rate of 1.9\%
yr$^{-1}$ from regions along the remnant's western limb.
We find that qualitatively, the observed spectral steepening and decline in
flux can be explained by a simple model for changes in the electron cutoff energy which
are brought about by a natural deceleration of the shock. We estimate an average
deceleration of Cas A's forward shock velocity $\approx$ 10 -- 40 km s$^{-1}$
yr$^{-1}$. 

The predicted decline in the nonthermal X-ray
emission is about 4\% yr$^{-1}$, which is nearly twice that observed.
The difference between the predicted and observed decline might be
explained by the fact that the 4.2--6.0 keV continuum emission is not entirely
due to synchrotron emission from shock accelerated electrons, but some of it is
from thermal continuum which does not evolve on the same timescale as the
nonthermal emission. We have compared our results to models where the
decline is a natural consequence of either a decrease in the number of
particles entering the shock or a decrease in the efficiency of the shock to
amplify the magnetic field, and find that these models predict a decline of
$\sim$ 0.1--0.5\% yr$^{-1}$, which is significantly less than the observed
decline of 1.5--1.9\% yr$^{-1}$. 

\acknowledgements

We would like to thank Alexey Vikhlinin and Paul Plucinsky for useful discussions regarding the
stability of the {\it Chandra} ACIS-S3 quantum efficiency.
D.~J.~P. acknowledges support from the {\it Chandra} GO program through
grant GO0-11094X as well as support from NASA contract NAS8-03060. J.~V. 
is supported by a Vidi grant from the Netherlands Scientific Organization (NWO).

\begin{deluxetable}{lcccccc}
\tablecolumns{7}
\tablewidth{0pc}
\tablecaption{Chandra Observations of Cas A and Spectral Fitting Results}
\tablehead{
\colhead{Epoch} & \colhead{$\Gamma_{\mathrm{SNR}}$} & \colhead{F\tablenotemark{a}} & 
\colhead{$\Gamma_{\mathrm{FS}}$} & \colhead{$\Gamma_{\mathrm{Interior}}$} & 
\colhead{$\Gamma_{\mathrm{West}}$}  & \colhead{F\tablenotemark{a}$_{{\mathrm{West}}}$}\\
\colhead{yr} & \colhead{} & \colhead{10$^{-10}$ erg cm$^{-2}$ s$^{-1}$} & 
\colhead{} & \colhead{} & \colhead{} & 
\colhead{10$^{-10}$ erg cm$^{-2}$ s$^{-1}$}}
\startdata
2000.1 & 2.81$\pm$0.03 & 1.61$\pm$0.01 & 2.32$\pm$0.11 & 2.66$\pm$0.07 & 2.66$\pm$0.06 & 0.229$\pm$0.001 \\
2002.1 & 3.01$\pm$0.03 & 1.56$\pm$0.01 & 2.43$\pm$0.11 & 2.75$\pm$0.07 & 2.74$\pm$0.06 & 0.223$\pm$0.001 \\
2004.1 & 2.99$\pm$0.03 & 1.54$\pm$0.01 & 2.42$\pm$0.11 & 2.70$\pm$0.07 & 2.73$\pm$0.06 & 0.215$\pm$0.001 \\
2007.9 & 2.98$\pm$0.04 & 1.45$\pm$0.02 & 2.55$\pm$0.15 & 2.70$\pm$0.10 & 2.80$\pm$0.09 & 0.197$\pm$0.002\\
2009.8 & 2.99$\pm$0.05 & 1.42$\pm$0.04 & 2.61$\pm$0.15 & 2.78$\pm$0.11 & 2.78$\pm$0.09 & 0.195$\pm$0.004 \\
2010.8 & 3.07$\pm$0.04 & 1.34$\pm$0.01 & 2.56$\pm$0.14 & 2.82$\pm$0.13 & 2.85$\pm$0.08 & 0.183$\pm$0.002
\enddata
\tablenotetext{a}{4.2 -- 6.0 keV flux}
\label{tab:data}
\end{deluxetable}

\end{document}